\pdfoutput=1
\documentclass[%
twocolumn,
superscriptaddress,
amsmath,
amssymb, 
aps,
pra,
floatfix,
nolongbibliography
]{revtex4-2}

\usepackage{graphicx}
\usepackage{bm}
\usepackage{amsmath}
\usepackage[dvipsnames]{xcolor}
\usepackage{hyperref}
\usepackage{physics}

\begin{document}

\title{Near-deterministic loading of molecular arrays by microwave-assisted collisions}

\author{Etienne F. Walraven}
\affiliation{Institute for Molecules and Materials, Radboud University, 6525 AJ Nijmegen, The Netherlands}
\author{Kang Feng}
\affiliation{Institute for Molecules and Materials, Radboud University, 6525 AJ Nijmegen, The Netherlands}
\author{Jonas Rodewald}
\affiliation{Centre for Cold Matter, Blackett Laboratory, Imperial College London, London SW7 2AZ, United Kingdom}
\author{Michael R. Tarbutt}
\affiliation{Centre for Cold Matter, Blackett Laboratory, Imperial College London, London SW7 2AZ, United Kingdom}
\author{Tijs Karman}
\email{t.karman@science.ru.nl}
\affiliation{Institute for Molecules and Materials, Radboud University, 6525 AJ Nijmegen, The Netherlands}
\date{\today}

\begin{abstract}
Molecular tweezer arrays offer great prospects for quantum simulation, sensing, and computing, and would benefit from methods that enhance loading efficiency.
Whereas light-assisted collisions underpin enhanced loading methods for atomic tweezer arrays, this approach cannot be directly extended to molecular arrays due to collisional loss.
We show how this collisional loss can be suppressed by shelving molecules in rotationally or vibrationally excited states, so that a shelved molecule interacts with a 
newly loaded molecule through a repulsive van der Waals interaction.
By introducing microwave assisted collisions, we show how to control the final states and the energy released in a collision between a pair of molecules. Following this controlled collision, one of the two molecules can be ejected, and we explore several strategies for ensuring deterministic ejection. Our schemes rely on currently available techniques for laser-coolable molecules, and we predict achievable filling fractions up to 95\%, paving the way for scalable molecular arrays.
\end{abstract}

\maketitle

\section{Introduction\label{sec:introduction}}

Arrays of neutral atoms and molecules using optical tweezer arrays have become a central platform in quantum information, quantum simulation, and quantum metrology \cite{kaufman:21}. Across these fields, there is an ever-increasing demand for scalable arrays, enabling higher stability optical clocks \cite{ludlow:15}, simulation of more complex collective phenomena \cite{ebadi:21}, and possible quantum advantage in quantum computing \cite{daley:22}. Currently, arrays may contain thousands of atoms \cite{gyger:24,norcia:24,pichard:24,lin:25,chiu:25,manetsch:25}, and tens of molecules  \cite{anderegg:19,holland:23,bao:23,vilas:24}. For most applications, these arrays need to be free of defects. However, conventional tweezer loading relies on collisional blockade, leading to arrays that are stochastically filled with a probability of about 50\% for atoms \cite{schlosser:01,schlosser:02} and 30--40\% for molecules \cite{anderegg:19, bao:23, holland:23,  vilas:24}. 
While rearrangement of tweezers can produce defect-free arrays~\cite{barredo:2016, manetsch:25, tian:23, cicali, sheng2021efficient},
the required rearrangement effort scales with array size, ultimately limiting the size of the array.

For an array of $N$ tweezers loaded independently with probability $p$,
the average number of defects is $N(1-p)$. The fidelity of transporting an atom into the desired location, $f$, is limited by imperfect identification of the site occupancies, loss due to collisions with background gas, loss due to photon scattering during imaging, and loss in the transport itself~\cite{pagano}. The fidelity of forming a defect-free array by rearrangement is $F=f^{N(1-p)}$ \cite{sheng2021efficient,harle2026atommovr}.
For example, when $f=0.999$, the fidelity of performing 1000 rearrangements is 36~\% but this falls to 0.005~\% for 10\,000 rearrangements.
The maximum array size that can be assembled with $F$ above some threshold scales as $(1-p)^{-1}$, so there is great advantage in increasing the initial loading probability~\cite{brown:19}. This is especially true for molecules where stochastic loading tends only to reach $p\approx 0.35$ and because $f$ is further limited by loss due to absorption of blackbody photons. 

Several methods have so far improved tweezer loading of atoms beyond the conventional 50\% filling fraction. Dark-state enhanced loading is an iterative approach based on imaging the loaded atoms, shelving them into long-lived states that are dark to the loading dynamics, and lowering the trap depths of the loaded tweezers to inhibit further loading \cite{shaw:23}. This approach has achieved a filling fraction of 84\%. Similarly, a recent paper proposes to use repulsive barriers to protect loaded atoms from collisions \cite{baldock:26}. Continuous operation schemes, on the other hand, use reservoirs to constantly refill the main atom array \cite{pause:23,gyger:24,norcia:24,chiu:25}, achieving the highest filling fractions of up to 99\%, yet requiring infrastructure for continuous rearrangement of tweezers. Lastly, light-assisted collisions have been used to eject all but one particle from each tweezer deterministically \cite{grunzweig:10,fuhrmanek:12,sompet:13,fung:15,lester:15,brown:19,aliyu:21,angonga:22,jenkins:22,pampel:25,muzi:25}. In such collisions, the detuning of the light provides a controlled energy release that can be tuned such that the probability for a single atom to escape the tweezer is maximized. 
Using such schemes, the typical loading efficiency for atoms is around 80\% \cite{brown:19},
while the highest reported array-averaged single-site occupancy is 93\% \cite{jenkins:22}. 

\begin{figure*}
\centering
\includegraphics[width=0.8\textwidth]{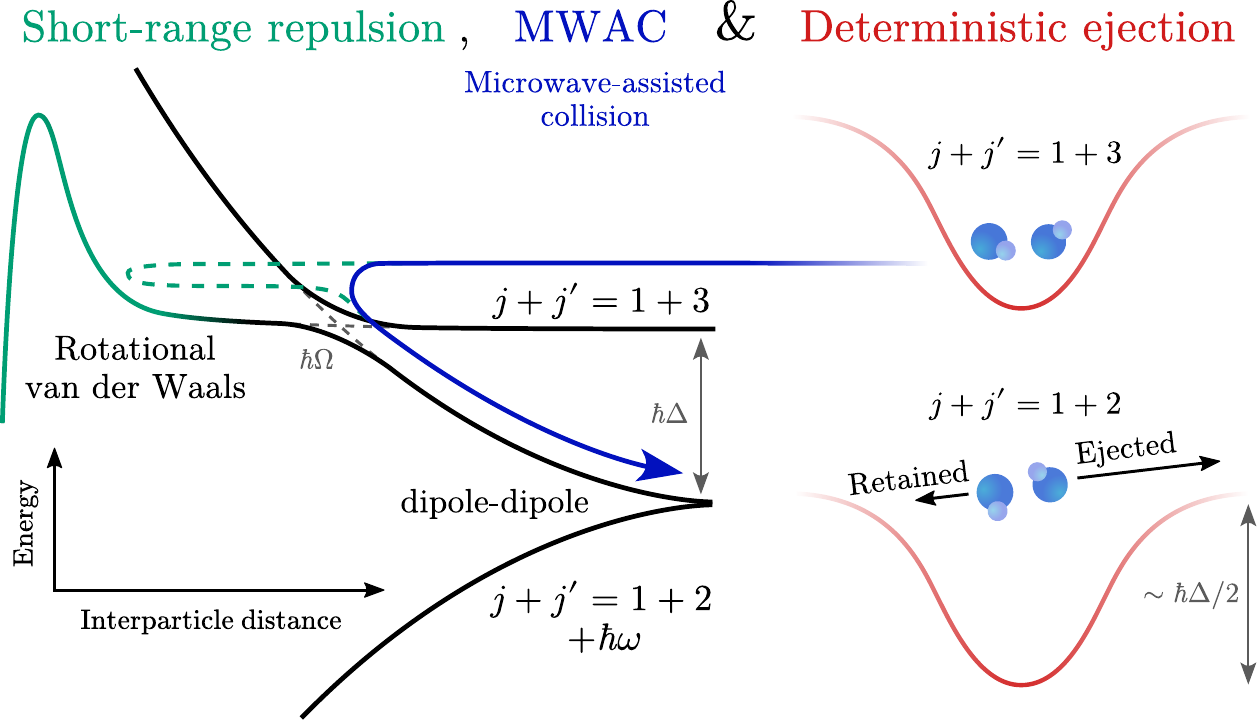}
\caption{
{\bf Sketch of a microwave-assisted collision with short-range repulsion and subsequent deterministic ejection.} Molecules in rotational states $j+j'=1+3$ experience repulsive rotational van der Waals interactions, reducing short-range loss. Dressing the molecule in $j=3$ to $j=2$ using microwaves creates resonant dipolar interactions between molecules in $j=1$ and the dressed state. In a microwave-dressed picture, as sketched here, the coupling due to the Rabi frequency $\Omega$ enables the inelastic process as shown in blue, releasing energy set by the detuning $\Delta$. Processes resulting in residual loss from tunneling to the short range or from transitions to other hyperfine states are not illustrated. The post-collision energy release then enables the ejection of a single particle, either using an asymmetry in momenta from the initial temperature, by utilizing the various molecular tensor Stark shifts, or by actively pushing out one of the two molecules.
\label{fig:sketch}
}
\end{figure*}

Polar molecules are a more recent addition to the field, with experiments already showing long coherence times in rotational states \cite{gregory:24, ruttley:25}, as well as the implementation of quantum gates based on rotational qubits \cite{holland:23, bao:23, picard:25, ruttley:25}. For laser-cooled molecules, stochastic loading probabilities have been observed between 30 and 40\% \cite{anderegg:19, bao:23, holland:23,  vilas:24}, while enhanced methods have not been implemented yet. The iterative feedback-based schemes of Refs.~\cite{shaw:23,baldock:26} could also be applied to molecules. In earlier work, we proposed an iterative scheme to deterministically load laser-cooled molecules into tweezers with an efficiency up to 80\%, by exploiting a dipolar blockade mechanism and collisional stability derived from a repulsive van der Waals interaction \cite{walraven:24b}. Here, we propose alternative enhanced loading methods for molecules inspired by the light-assisted collision schemes for atoms.

The core principle of atomic light-assisted collisions relies on non-adiabatic transitions during the collision between two atoms, where one atom is excited to a higher-lying electronic state by blue-detuned cooling light \cite{bali:94}.
After excitation the particles gain kinetic energy that is at most the detuning,
but due to spontaneous emission they are more likely to gain a smaller fraction~\cite{sompet:13,pampel:25}.
This energy release is not sharply defined, but rather stochastic due to the role of spontaneous emission.
For atoms at rest, this energy is shared equally between the atoms due to momentum conservation, but a momentum imbalance can arise from a non-zero initial temperature, so there is potential for just one particle to escape \cite{grunzweig:10,sompet:13}. If both atoms remain trapped, subsequent elastic collisions can redistribute the momentum, and there can be multiple light-assisted collisions. Consequently, light-assisted collisions can eject all but a single atom, thereby loading tweezers deterministically with single atoms. For ultracold molecules, this concept cannot be applied directly as short-range encounters typically result in universal loss rather than elastic collisions \cite{takekoshi:14,voges:20,guo:18,ye:18,gregory:19,cheuk:20}. This universal loss prevents enhanced loading \cite{walraven:24b}. While microwave shielding could in principle create repulsive interactions that decrease this loss \cite{karman:18,lassabliere:18,anderegg:21}, it disrupts laser cooling by dressing the $j=1$ rotational state with another rotational state for which the cooling is not rotationally closed. 

In this paper, we propose several strategies for the deterministic loading of optical tweezer arrays with laser-cooled molecules. They rely on three main components, as illustrated in Fig.~\ref{fig:sketch}: shelving into states with repulsive interactions, controlled collisional energy release and deterministic single-molecule ejection. 
The collisional loss that prevents direct application of light-assisted collisions is suppressed by exploiting a repulsive van der Waals interaction that arises when the colliding molecules are in different rotational or ro-vibrational states. 
The controlled energy release can then be realized by dressing rotational states with microwaves, enabling a new type of collision process that we term microwave-assisted collisions (MWACs), analogous to light-assisted collisions (LACs) for atoms.
We describe how this engineered interaction leads to controlled energy release, whose efficiency we quantify through coupled-channels scattering calculations. Following this controlled collision, a single molecule can be ejected deterministically. We consider several single-molecule ejection methods: thermal ejection, push-beam ejection, trap-lowering ejection, and tensor-Stark ejection. 
Based on these ideas, we present experimentally viable loading schemes, which iteratively increase the filling fraction of a molecular tweezer array. 
When rotational van der Waals repulsion is utilized, we find that filling fractions of up to 87\% can be achieved, limited by residual collisional loss. Using the stronger ro-vibrational van der Waals repulsion eliminates this collisional loss, leading to filling fractions up to 95\%, limited by the lifetime of the vibrationally excited state. 

\section{Concept}\label{sec:concept}

Our aim is to control the energy released in a collision between two tweezer-trapped molecules. To do that, the first challenge is to suppress the short-range encounters that produce uncontrolled, lossy collisions. We do this by engineering a repulsive van der Waals interaction that prevents the molecules from reaching short range. Consider a pair of molecules in rotational states $j,j'$ and label this pair state $j+j'$. Its energy is $E_{j,j'}=B[j(j+1)+j'(j'+1)]$. The van der Waals interaction is the second-order dipole-dipole coupling to other pair states. It is necessary to sum over all relevant pair states, but this sum may be dominated by a single term when one pair state, $k+k'$, is much closer in energy than all the others. In this case the interaction is proportional to $|\bra{k}\hat{d}\ket{j}|^2\,|\bra{k'}\hat{d}\ket{j'}|^2/(R^6\Delta E)$, where $\hat{d}$ is the dipole operator and $\Delta E = E_{j,j'}-E_{k,k'}$ is the energy difference between the interacting pair states. When $\Delta E$ is small and positive the interaction will be strong and repulsive. When $|j'-j| e 2$, the interaction is always repulsive~\cite{walraven:24a}. Taking $j+j'=1+3$ as an example, the closest pair state is $2+2$, which lies lower in energy by $2B$ resulting in repulsion at long range that is strong enough to effectively suppress short-range encounters. This repulsive interaction is ubiquitous for polar molecules where the rotational contribution to the van der Waals interaction far outweighs the electronic one. The repulsion can be made even stronger by using different ro-vibrational states. To see this, consider the pair of states $(v,j)+(v',j') = (0,1)+(1,0)$. The dipole-dipole interaction couples it to $(0,0)+(1,1)$ which is slightly lower in energy due to the small difference in the rotational constants of $v=0$ and $v=1$, giving $\Delta E = 2(B_{v=0}-B_{v=1})$. Since $\Delta E \ll B$, this repulsive ro-vibrational van der Waals interaction \cite{feng:2026} is much stronger than the repulsive rotational van der Waals interaction discussed above.

Having suppressed the unwanted short-range collisions, we now introduce a controlled microwave-assisted collision (MWAC) similar to the light-assisted collisions used for atoms. Figure~\ref{fig:sketch} illustrates the idea for the case where the initial state is $j+j'=1+3$, so that there is a repulsive rotational van der Waals interaction. A microwave field red detuned by $\Delta$ from the $j'=3\rightarrow 2$ transition brings $1+2$ close to $1+3$ in a dressed-state picture. The pair state $1+2$ has a resonant dipole-dipole interaction whose repulsive branch crosses $1+3$ at an interparticle distance set by $\Delta$, and the microwave coupling turns this into an avoided crossing. As the molecules approach and separate, they traverse the crossing twice, and if one traversal is diabatic and the other adiabatic they will be transferred to $1+2$, gaining energy $\hbar\Delta$. This energy release can then be used to to eject one of the two particles with high probability.

With these foundations, we can sketch the complete iterative scheme for deterministic loading of a tweezer.
At the beginning of each cycle, a tweezer is either empty or contains a single shelved molecule (in $j=3$ or $(v,j)=(2,0)$, depending on the chosen scheme).
Then, each cycle consists of the following steps:
\begin{enumerate}
    \item {\bf Load} -- Load molecules in $j=1$ by laser cooling.
    \item {\bf MWAC} -- Apply a microwave field to induce a collision with a precise, chosen kinetic energy release.
    \item {\bf Eject} -- Allow one of the two molecules to escape, or actively drive it out.
    \item {\bf Shelve} -- Transfer the molecule to a state that has a repulsive van der Waals interaction with $j=1$. 
    \item {\bf Iterate} -- Repeat from step 1.
\end{enumerate}

\begin{figure*}
\centering
\includegraphics[width=0.8\linewidth]{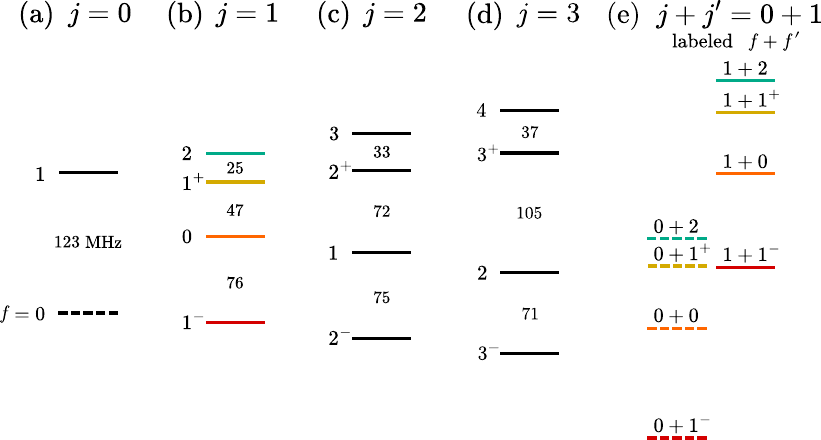}
\caption{
{\bf CaF hyperfine structure} in rotational states (a) $j=0$, (b) $j=1$, (c) $j=2$, and (d) $j=3$.
Panel (e) shows the hyperfine structure of a \emph{pair} of molecules in $j+j'=0+1$. Here, color and line style encode hyperfine states in $j=0$ and $j=1$, respectively. The hyperfine intervals are shown in MHz.
\label{fig:energylevels}
}
\end{figure*}

In the following sections, we describe each of these steps in detail. To give a concrete realization, we explain how the steps work for CaF molecules, since tweezer experiments with this species are already well advanced~\cite{anderegg:19, cheuk:20, holland:23, bao:23, lu:24, holland:25, lu:26, holland:26}. The same ideas will also work for other laser-cooled molecules. We make extensive use of the hyperfine components of the low-lying rotational states. For CaF, these are shown in Fig.~\ref{fig:energylevels}.

\section{Load}\label{sec:load}

Molecules in rotational state $j=1$ are loaded into optical tweezers using standard laser cooling methods~\cite{Fitch:21}. Several gray molasses cooling schemes are available that reach temperatures of a few $\mu$K, both in free space and in optical traps \cite{caldwell:19, burau:24}. These cooling schemes take about 1~ms to reach the equilibrium temperature. The loading time should be long compared to this so that loaded molecules are likely to be cold, but short enough that the probability of loading more than one molecule into a tweezer is small.
To infer typical loading times we note that tweezers are usually loaded in the collisionally blockaded regime, which is effective for loading rates of 1--100 particles per second~\cite{schlosser:01,schlosser:02}.
Through the choice of density, we set the loading rate to about 20 molecules per second, and choose a loading time of 5~ms, so that the average loading probability is about 0.1 per tweezer in each loading cycle.

CaF, and similar laser-coolable molecules, have four hyperfine components within $j=1$, arising from spin-rotation and electron-nuclear spin interactions. They are $f=0,1^-,1^+,2$ where $f$ is the total angular momentum quantum number and the two states with $f=1$ are distinguished by the $\pm$ superscript, with $1^-$ having the lower energy.
This hyperfine structure is illustrated in Fig.~\ref{fig:energylevels}(b). 
Laser cooling tends to leave molecules distributed amongst the 12 Zeeman sub-levels of these four hyperfine states, but the subsequent steps work best for molecules in a single quantum state. We suggest using the deep cooling method described in \cite{caldwell:19}, because it reaches temperatures around 5~$\mu$K in free space and drives the molecules towards a dark state within $f=2$ only. We assume that a similar temperature will be reached in the tweezer since the dark state is robust to the tensor Stark shifts produced by the tweezer light. If necessary, the cooling step can be followed by a brief optical pumping pulse that prepares molecules in a single state. In particular, the state $(f,m_f)=(2,2)$ [or $(2,-2)$] can be prepared with high fidelity~\cite{holland:25}.

\section{Microwave-assisted collisions\label{sec:mwac}}

To help understand microwave-assisted collisions (MWACs), we first review the light-assisted collisions used for atoms. At long range, the interaction between two ground state atoms is an attractive van der Waals interaction. The interaction between a ground state atom and an excited atom is a resonant dipole-dipole interaction with an attractive branch and a repulsive branch. 
In a dressed-state picture, for positive, blue detuning, the van der Waals state crosses the repulsive dipole-dipole state at the so-called Condon point.
The coupling of the states by the light turns this into an avoided crossing.
If the approaching particles traverse the crossing diabatically, they will have a short-range encounter. If they traverse the crossing adiabatically they will repel and separate, traversing the crossing again on their way out. If this second crossing is diabatic, the particles will gain kinetic energy equal to the detuning (or some fraction of it, see below). This is a light-assisted collision with a controlled energy release. However, the probability of a short-range encounter is always greater than that of the light-assisted collision \cite{walraven:24b}. For atoms, this is not an issue, as short-range collisions lead to redistribution of momentum. For molecules, on the other hand, this leads to universal loss, and for such an approach to work, we require the interaction between the two molecules to be repulsive at shorter separation than the Condon point.

Instead of using light to couple electronic states, we use microwaves to couple rotational states, as illustrated in Fig.~\ref{fig:sketch} and described in Sec.~\ref{sec:concept}. This has two major advantages: we can utilize the repulsive van der Waals interactions described in Sec.~\ref{sec:concept}, and the rotational states have such long lifetimes that spontaneous emission is negligible.  In contrast to atomic light-assisted collisions, where a non-deterministic fraction of the detuning is released as kinetic energy by fast spontaneous emission \cite{pampel:25}, MWACs yield the precisely detuning as the energy release. It follows that a molecule can be ejected through a single collision as the energy release is well defined. We explore this in Sec.~\ref{sec:ejection}. In this section, we consider the MWAC process for molecule pairs in the same vibrational state (utilizing rotational van der Waals repulsion) and for molecules in different vibrational states (utilizing ro-vibrational van der Waals repulsion).

\subsection{Calculation methods}

We calculate collision rates through coupled-channels scattering calculations using the renormalized Numerov method \cite{johnson:78}, with a short-range absorbing boundary condition and long-range $S$-matrix boundary conditions \cite{janssen:13}. We treat the molecules as rigid rotors with hyperfine structure, tensor Stark shifts and microwave-dressed states, interacting via electrostatic interactions through the dipole and quadrupole moments, including the electronic van der Waals interaction, as described in more detail in Refs. \cite{walraven:24a,walraven:25,karman:18}. 
Molecular vibrations are included as described in Ref.~\cite{feng:2026}.
We then define the collisional MWAC efficiency $P_\mathrm{MWAC}$ using collision rate coefficients $k$ of the various processes as
\begin{equation}
    P_\mathrm{MWAC}=\frac{k_\mathrm{MWAC}}{k_\mathrm{MWAC}+k_\mathrm{short}+k_\mathrm{inel}}\,,
\end{equation}
where `short' stands for short-range inelastic collisions and `inel' stands for all state-changing inelastic loss processes, excluding the inelastic rates to desired microwave-dressed channels. The residual loss consists of short-range loss by tunneling through the barrier set by the repulsive van der Waals interaction, as well as inelastic loss to undesired channels, especially nearby hyperfine states. 

\subsection{MWAC using purely rotational states}

\begin{figure}
\centering
\includegraphics[width=\linewidth]{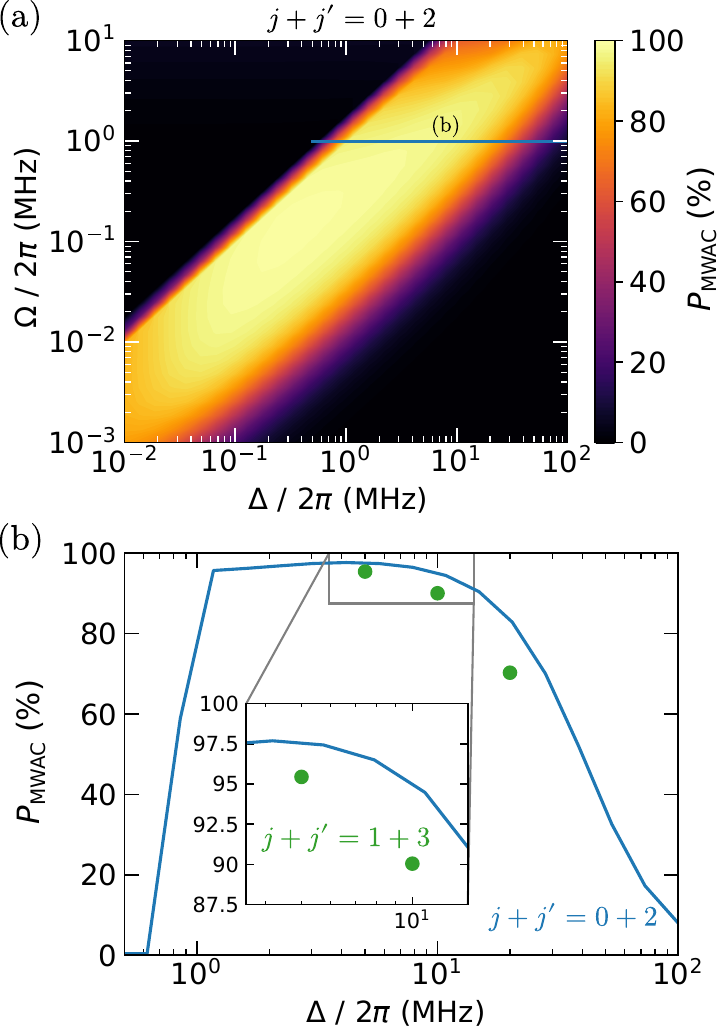}
\caption{
{\bf Microwave-assisted collision (MWAC) efficiency}. (a) Efficiency as a function of Rabi frequency $\Omega$ and detuning $\Delta$ for $j+j'=0+2$ with $m=m'=0$. We have assumed a temperature of 5~$\mu$K  and a microwave field on the transition $j=2\to 1$. These results do not account for hyperfine structure and tensor Stark shifts. (b) Efficiency for $\Omega=2\pi\times1$~MHz, where the blue curve represents the data from (a), as indicated by the blue line. The data points in green are for $j+j'=1+3$ in hyperfine states $(j,f,m_f)+(j',f',m_f')=(1,2,1)+(3,4,4)$, with a microwave field on $(j',f',m_f')=(3,4,4)\to(2,3,3)$ and tensor Stark shifts set by tweezer depths of $\hbar\Delta/2$.
\label{fig:collisions}
}
\end{figure}

The MWAC efficiency, $P_\mathrm{MWAC},$ is defined above as the branching of collision rates to the desired energy release. This efficiency depends on the detuning $\Delta$ and { the} Rabi frequency $\Omega$ of the microwave field. To build intuition, we first explore this parameter space for CaF molecules in $(j,m_j)=(0,0)$ and $(2,0)$ with a microwave field addressing $j=2\to1$. This is the simplest case for which MWACs can be engineered, which is also computationally the least expensive, while it preserves the essential features. For simplicity, we neglect hyperfine structure and tensor Stark shifts due to the tweezer in this case, with results converged to $L=7$ partial waves at a temperature of 5~$\mu$K. Figure~\ref{fig:collisions}(a) shows the calculated MWAC efficiency as a function of $\Delta$ and $\Omega$. We see that the efficiency can be high for a very wide range of $\Omega$ and $\Delta$ values, provided $\Omega$ is between $0.01\Delta$ and $\Delta$. The optimum occurs when $\Omega\sim0.1\Delta$. As we discuss in Sec.~\ref{sec:ejection}, a detuning of about twice the trap depth enables efficient ejection of molecules. Efficient loading typically { uses a trap depth of at least $V_0/h=5$~MHz~\cite{brooks:2021, anderegg:19}}, so a realistic lower bound for the detuning is {$\Delta\sim 2\pi\times10$~MHz}, calling for {$\Omega \sim 2\pi\times1$~MHz}. The blue line in Fig.~\ref{fig:collisions}(b) shows the efficiency versus $\Delta$ at {$\Omega = 2\pi\times1$~MHz}. It reaches 97.5\% at {$\Delta=2\pi\times4$~MHz}, falling to 95\% at {$\Delta=2\pi\times10$~MHz} and dropping below 90\% for {$\Delta>2\pi\times20$~MHz}.

Since laser cooling leaves molecules in $j=1$, we next consider the case where one molecule is in $j=1$ and the other is shelved in $j=3$, as sketched in Fig.~\ref{fig:sketch}. We engineer MWACs by applying microwaves on the $j'=3\to2$ transition. This dressing does not interfere with the laser cooling of the molecules in $j=1$, which means that the dressing can be applied continuously.

While MWACs can work for any of the hyperfine states $(f,m_f)$, it works better for some states than others, so we need to consider the optimum choice. First, we note that the tensor Stark shift causes molecules in different $(f,m_f)$ to experience different tweezer depths. For the energy release after MWAC to be as well-defined as possible,
{ we initialize the $j=3$ molecules into a specific ($f,m_f$) state.}
Figure~\ref{fig:energylevels}(d) shows these states. A suitable choice is $(j,f,m_f)=(3,4,4)$. 
We provide example schemes of shelving into this state in Sec.~\ref{sec:shelve}.
The microwave field needed for MWAC is then { a single frequency field} tuned close to resonance with the transition $(j,f,m_f)=(3,4,4)\to(2,3,3)$.

\begin{figure}
\centering
\includegraphics[width=\linewidth]{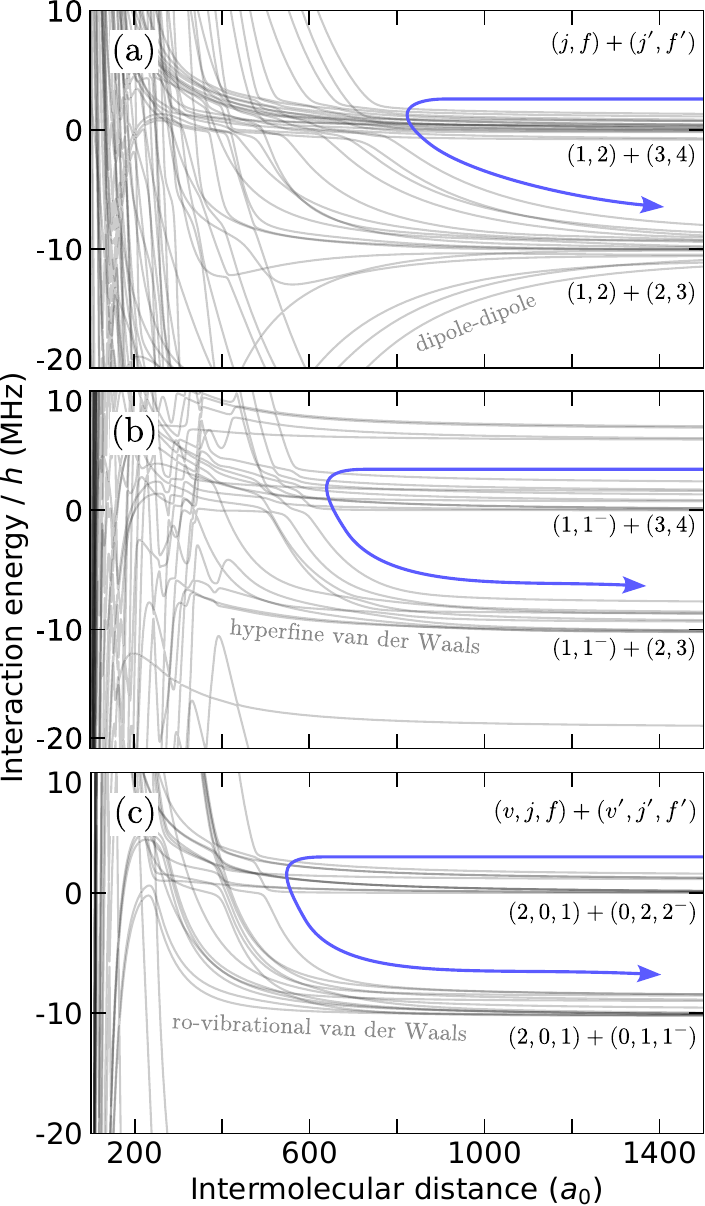}
\caption{
{\bf Microwave-dressed CaF interaction potentials.} (a,b) Molecules are in different rotational states with microwave dressing on $(j,f)=(3,4)\to(2,3)$. (c) Molecules are in different ro-vibrational states with microwave dressing on $(v,j,f)=(0,2,2^-)\to(0,1,1^-)$. The blue arrows indicate the MWAC trajectories. The three panels show the scenarios where the post-MWAC states experience (a) dipole-dipole, (b) hyperfine van der Waals and (c) ro-vibrational van der Waals interactions. Only $L=0$ and 2 are shown here with (a,b) $M_\mathrm{tot}=0$ and (c) $M_\mathrm{tot}=3$.
\label{fig:potentials}
}
\end{figure}

Figure~\ref{fig:potentials} shows the adiabatic interaction potentials obtained by diagonalization of the full Hamiltonian, except radial kinetic energy, at every intermolecular distance $R$. In Fig.~\ref{fig:potentials}(a), we focus on the case $(j,f)+(j',f')=(1,2)+(3,4)$. 
The microwave dressing is on the transition $(j,f)=(3,4)\to(2,3)$.
Because the interaction between $(j,f)=(1,2)$ and $(2,3)$ is dipolar, scaling as $R^{-3}$, the Condon points with $(1,2)+(3,4)$ are formed at large distances. For the other hyperfine states of $j=1$ ($f=1^-,0,1^+$), dipolar selection rules prevent first-order coupling, leading to repulsive $R^{-6}$ hyperfine van der Waals interactions \cite{walraven:25}. This interaction is stronger than the rotational van der Waals, so we still obtain Condon points but at shorter distances than for the dipolar case,
as seen in Fig.~\ref{fig:potentials}(b) for $j=1,f=1^-$.
Consequently, the range of detunings where MWAC is effective is smaller when $f\neq2$. Nevertheless, for our particular choice of detuning {($\Delta=2\pi\times10$~MHz)}, we find that the collisional loss and MWAC rates are comparable amongst all $f$ for the $(1,f)+(3,4)$ states so that MWAC can be effective in all cases. { This means that the scheme is not dependent on the fidelity with which the deep laser cooling scheme drives the molecules into the $f=2$ dark state.} 
In summary, we propose to use the following CaF hyperfine states: $(j,f,m_f)=(1,2,\pm1)$ from deep laser cooling, $(3,4,4)$ as the shelved state, with microwaves red detuned from $(2,3,3)$.

Using the methods described above, we calculate collision rate coefficients for one molecule in $(j,f,m_f)=(1,2,\pm1)$ and the other in in $(3,4,4)$,  with a microwave field close to the $(3,4,4)\to (2,3,3)$ transition, using a basis set up to $L=5$. Here, we include hyperfine structure and tensor Stark shifts. We take a tweezer depth of $V_0=\hbar\Delta/2$, a scalar polarizability of  $\alpha_0=1.4\times10^{-3}$~Hz/(W/m$^2$) and a tensor polarizability of $\alpha_2=-0.8\times10^{-3}$ Hz/(W/m$^2$), which are the appropriate values for a CaF molecule at 780~nm. The green points in Fig.~\ref{fig:collisions}(b) show the results at { $\Omega=2\pi \times 1$~MHz}, for three different values of $\Delta$. The MWAC efficiency is 95.5\% at {$\Delta = 2\pi\times5$~MHz}, and 90.0\% at {$\Delta=2\pi\times 10$~MHz}. We also see from Fig.~\ref{fig:collisions}(b) that the results for this case are not too far from the results obtained for the simpler $j+j'=0+2$ system where hyperfine structure and tensor Stark shifts were neglected, 
indicating that the simpler calculation can be a useful qualitative guide.
To quantitatively determine the MWAC efficiency, however, we use the results of the complete calculation.

\subsection{MWAC using ro-vibrational states}

Next, we consider the case where the molecules are shelved in a different rotation-vibration state $(v,j)=(2,0)$. This has the advantage of almost eliminating the residual collisional loss because the repulsive ro-vibrational van der Waals interaction \cite{feng:2026} between a $(v,j)=(2,0)$ molecule and a $(v,j)=(0,1)$ molecule is far more effective than the rotational van der Waals interaction. The scheme is similar to the one sketched in Fig.~\ref{fig:sketch}: the new molecule loaded in $(v,j)=(0,1)$ is first transferred to $(0,2)$ using a short microwave pulse, and then the same microwave field induces a MWAC by dressing between $(2,0)+(0,1)$ (ro-vibrational van der Waals) and $(2,0)+(0,2)$ (rotational van der Waals). This produces molecules in the pair state $(2,0)+(0,1)$ with a tuneable energy release. To illustrate this, we show adiabatic interaction potentials using these ro-vibrational states in Fig.~\ref{fig:potentials}(c). Since the ro-vibrational van der Waals is stronger than the rotational one, Condon points are formed between the prepared and MWAC channels.
After the MWAC, one of the molecules is ejected, see Sec.~\ref{sec:ejection}.
Microwave pulses transfer the retained molecule to $(2,1)$ or $(0,1)$, followed by re-cooling (the same cooling light works for both cases) and shelving in $(v,j)=(2,0)$, see Sec.~\ref{sec:shelve}. After each iteration, more tweezers will be filled with single molecules in $(v,j)=(2,0)$, but never more than one.

\begin{figure}
\centering
\includegraphics[width=\linewidth]{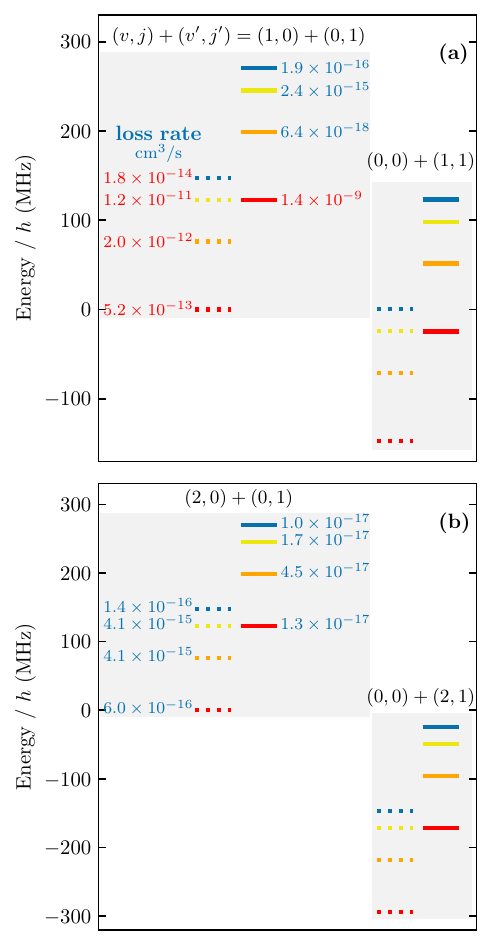}
\caption{
{\bf Loss rates for shelving in ro-vibrational states}.
Panels (a) and (b) correspond to shelving in $(v,j)=(1,0)$ and $(2,0)$, respectively.
\label{fig:vib_bgloss}
Dashed (solid) denote $f=0$ ($f=1$) of the $j=0$ molecule, whereas the colors encode $f=1^-,0,1^+,2$ of the $j=1$ molecule, see Fig.~\ref{fig:energylevels}.
Loss rates above $10^{-14}$~cm$^3$/s are highlighted in red, indicating loss is not suppressed sufficiently to completely eliminate { collisional loss of co-trapped molecules}.
}
\end{figure}

We quantify collisional loss rate coefficients by coupled-channels calculations as before,
but now including ro-vibrational energy level structure as explained in more detail in Ref.~\cite{feng:2026}.
These calculations include AC tensor Stark shifts and are performed at a temperature of 5~$\mu$K, although the loss rate coefficients are essentially temperature independent in this regime.
The resulting rates are shown per pair of rotation-vibration-hyperfine states in Fig.~\ref{fig:vib_bgloss}(a) and (b) for shelving in $v=1$ and $v=2$, respectively.
Loss rates can be suppressed below $10^{-14}$~cm$^3$/s, which almost eliminates collisional loss on all relevant timescales even for high in-tweezer densities exceeding $10^{14}$~cm$^{-3}$.
However, such low loss rates only occur for hyperfine states in $(v,j)+(v',j')=(v,0)+(0,1)$ that are higher in energy than all hyperfine states in the pair $(0,0)+(v,1)$.
For shelving in $v=1$, this is realized in many but not all hyperfine states, as can be seen in Fig.~\ref{fig:vib_bgloss}(a).
In this case, the $(0,0)+(1,1)$ channel is lower by $2\alpha_e = 147$~MHz which is comparable to the combined hyperfine structure in both molecules \cite{anderson:94}.
By shelving in $v=2$, the energy splitting is doubled, and collisional loss is effectively eliminated in every hyperfine state,
as can be seen in Fig.~\ref{fig:vib_bgloss}(b).
Therefore, we propose shelving in $v=2$ despite its shorter radiative lifetime (120~ms) relative to that of $v=1$ (240~ms).

\begin{figure}
\centering
\includegraphics[width=\linewidth]{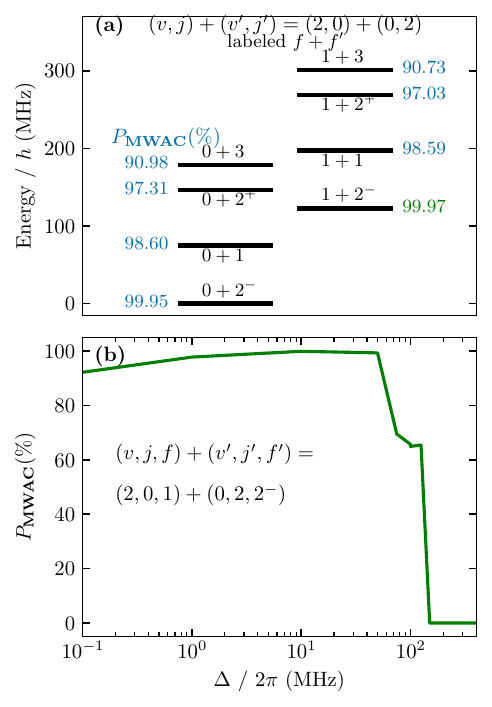}
\caption{
{\bf MWAC efficiency for ro-vibrational shelving}
(a) computed for detuning {$\Delta = 2\pi\times10$~MHz} and Rabi frequency {$\Omega = 2\pi\times1$~MHz} for various hyperfine states,
and (b) computed as a function of the detuning for a single pair of hyperfine states $(v,j,f)+(v',j',f')=(2,0,1)+(0,2,2^-)$.
In all cases, the tweezer depth is half the detuning.
\label{fig:vib_mwac}
}
\end{figure}

Figure~\ref{fig:vib_mwac} shows the MWAC efficiency for shelving in $v=2$,
defined as the fraction of collisions with the desired energy release equal to the detuning.
Panel~(a) shows the MWAC efficiency for various choices of hyperfine states. The fidelity exceeds 90\% for all choices and is 99\% or higher in several cases.
These states can be prepared by optical pumping followed by microwave transfer.
Unlike the purely rotational case considered previously, the MWAC efficiency of this ro-vibrational shelving scheme is so high that the fidelity of tweezer loading is no longer limited by collisions but instead by the lifetime of the vibrationally excited state.

Figure~\ref{fig:vib_mwac}(b) shows the dependence of the MWAC efficiency on the detuning.
When compared to the case of purely rotational shelving, Fig.~\ref{fig:collisions}(b),
it is apparent that the MWAC efficiency remains much closer to 100\% for much deeper tweezers.
In the purely rotational case, this is attributed to the barrier height in the $j+j'=1+2$ channel,
which is limited by dipole-dipole coupling to $j+j'=2+3$.
The detuning must be smaller than this maximum barrier height to form an effective Condon point, see Fig.~\ref{fig:potentials}.
In the ro-vibrational case, the barrier height is similarly limited by coupling of the $(v,j)+(v',j')=(1,0)+(0,1)$ channel to $(1,1)+(0,2)$,
but the resulting repulsive barrier is much higher.
In fact, in this case the maximum detuning is no longer limited by the barrier height,
but rather by dressing with another hyperfine state { that occurs when the detuning exceeds tens of MHz.}
The ability to perform efficient MWACs at much larger detunings suggests that this scheme could be effective for much deeper tweezers and higher temperature than the examples used in this paper (5~MHz depth and 5~$\mu$K temperature),
but we do not explore this further. 

\section{Ejection\label{sec:ejection}}

The MWAC releases kinetic energy, just like for light-assisted collisions in atomic enhanced loading. In the atomic case, it has been shown that the momentum asymmetry from the pre-collision temperature can lead to loss of a single atom after multiple collisions \cite{grunzweig:10,sompet:13}. For molecules, we would like to eject one of the two molecules after a single collision. In this section, we explore several strategies to achieve this and determine their efficiencies. The strategies apply to the final states produced by both the rotational and ro-vibrational schemes; where we need to be specific we will focus on the rotational scheme.

\subsection{Thermal ejection and the role of tensor Stark shifts}\label{sec:ejection_thermal}

By choosing a collisional energy release that does not exceed twice the trap depth, we create a situation where one molecule could escape, while the other remains trapped. 
The efficiency of single-molecule ejection from a trap depends on the trap depth and in-tweezer temperature prior to the collision. We consider molecules initially trapped in internal states $i$ in tweezers of depth $V_i=V_0(1+g_i)$ with trapping frequency $\omega_i=\omega_0 \sqrt{(1+g_i)}$, where $V_0$ and $\omega_0$ are the tweezer depth and frequency due to the isotropic part of the polarizability, and $g_i$ is the state-dependent modification due to the tensor Stark shift. After a MWAC, the molecules end up in final states $f_i$, releasing an amount of energy that is state-dependent through $\hbar\Delta=\hbar\Delta_0+V_0(g_{f_1}+g_{f_2}-g_{i_1}-g_{i_2})$, with $\Delta_0$ the detuning without tensor Stark shifts. 

\begin{figure}
\centering
\includegraphics[width=\linewidth]{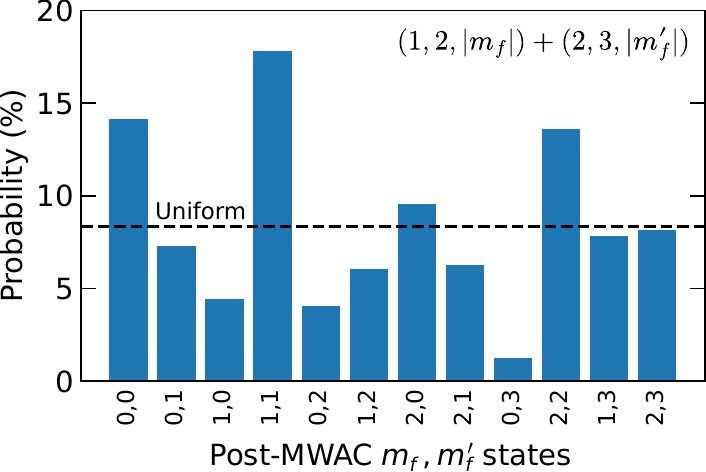}
\caption{
{\bf Distribution of post-MWAC states} of the various $m_f,m_f'$ product pairs of $(j,f,m_f)=(1,2,m_f)$ and $(2,3,m_f')$ after a MWAC between $(1,2,1)$ and $(3,4,4)$. These are obtained at {$\Delta=2\pi\times10$~MHz}, $V_0/h=5$~MHz, and $E/k_B=5$~$\mu$K from the initial $s$-wave channel $L=0$. The pairs are arranged from the largest positive to the largest negative tensor Stark shifts. For thermal ejection, we assume a uniform distribution as presented by the dashed line.
\label{fig:outgoingM}
}
\end{figure}

The MWAC products from a collision between our proposed hyperfine states $(j,f,m_f)=(1,2,1)$ and $(3,4,4)$ end up in the $(1,2,m_f)+(2,3,m_f')$ manifold. Which final $m_f,m_f'$ states dominate depends on the dynamics of the MWAC. In Fig.~\ref{fig:outgoingM}, we show the distribution of these $m_f,m_f'$ for {$\Delta=2\pi\times10$~MHz}, $V_0/h=5$~MHz, and $E/k_B=5$~$\mu$K from the initial $s$-wave channel. The final states experience slightly varying tweezer depths, and to capture the efficiency of ejection, we need to account for this product distribution by averaging over the various final states. In the following, we approximate the final state distribution as a uniform distribution, since there are no particularly dominant post-MWAC states.

We investigated the efficiency of deterministic ejection using a simple Monte Carlo (MC) simulation, which randomly samples pre-collision momenta $\bm{p}$ and center-of-mass positions $\bm{R}$ for both molecules, as well as the direction of the momenta imparted by the collision. Since $T\ll V_i/k_B$, we approximate the tweezer to be harmonic with {$\omega_x=\omega_y=2\pi\times50$~kHz} and {$\omega_z=2\pi\times5$~kHz}, drawing the positions and momenta from their corresponding thermal Gaussian distributions. We then update the relative momentum vector to point in a uniformly distributed random direction, effectively assuming an isotropic collision cross section, with a magnitude corresponding to energy conservation $p_f^2=p_i^2+2\mu\hbar\Delta(\bm{R})$. Here, the energy released is position-dependent through the local trapping potential as $\hbar\Delta(\bm{R})=\hbar\Delta_0+(V_0-\frac{1}{2}m[\omega_x^2X^2+\omega_y^2Y^2+\omega_z^2Z^2])(g_{f_1}+g_{f_2}-g_{i_1}-g_{i_2})$. This yields final kinetic energies of the two molecules, and if either exceeds the local trapping potential, we classify that molecule as ejected. This ejected molecule does not return for secondary collisions. 
This is in contrast to atomic enhanced loading, where due to spontaneous emission the energy release is typically a fraction of the detuning and not sharply-defined.
Therefore, the energy release is not necessarily sufficient to fully eject a single atom.
In that case, secondary light-assisted collisions take place that eventually eject an atom.

\begin{figure}
\centering
\includegraphics[width=\linewidth]{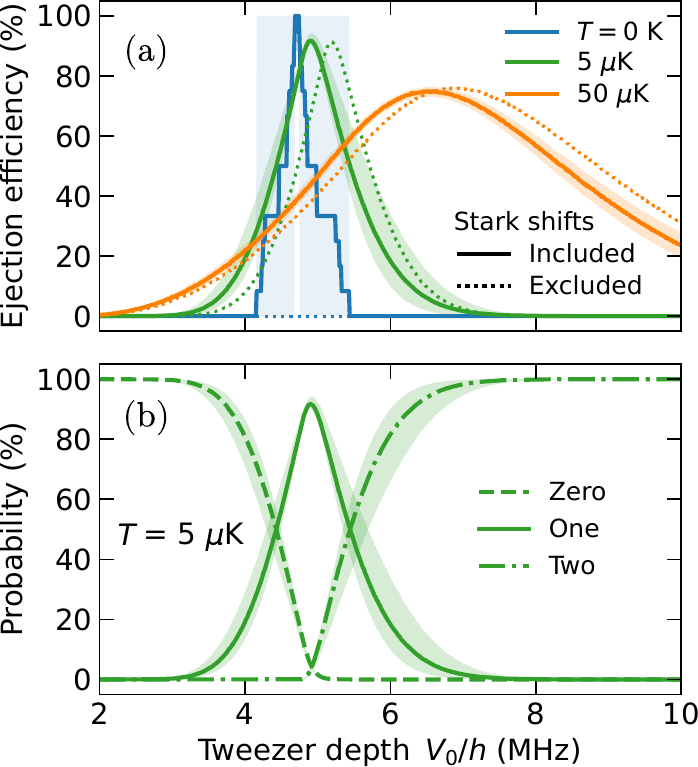}
\caption{
{\bf Thermal deterministic ejection efficiency} of molecules from an optical tweezer after a successful MWAC with {$\Delta=2\pi\times10$~MHz}. (a) Probability of ejection of a single molecule at $T=0,5$ and 50~$\mu$K, given initial states before the collision of $(j,f,m_f)=(1,2,1)$ and $(3,4,4)$. Solid curves represent a uniform average over all resulting $m_f,m_f'$ states of $(1,2,m_f)+(2,3,m_f')$, with shaded areas between the minima and maxima across all $m_f,m_f'$ states due to the state-specific tensor Stark shifts. The dotted curves neglect these Stark shifts. (b) Probabilities at $T=5$~$\mu$K of retaining zero, one, or two molecules.
\label{fig:ejection}
}
\end{figure}

Figure~\ref{fig:ejection} shows the ejection efficiency given our proposed $(j,f,m_f)=(1,2,1)$ and $(3,4,4)$ hyperfine states at a detuning of { $2\pi\times10$~MHz}, for $T=0,5$ and 50~$\mu$K. 
The range of efficiencies for the individual final $m_f,m_f'$ states is indicated by the shaded regions around the uniform average shown by the solid curves, where for each final state we took $10^5$ MC samples. For each state, we obtain $g_i$ as a function of $V_0$ by calculating their AC Stark shift. For example, at $V_0/h=10$~MHz, we find $g_{(1,2,1)}=-0.079$ and $g_{(3,4,4)}=0.19$.
The dotted curves represent the simplified scenario where we neglect the tensor Stark shifts, i.e., $g_i=0$, such that all states experience the same tweezer depth. 
In the low-temperature limit, $T=0$, energy and momentum conservation require the post-collision momenta to be in opposite directions and with equal magnitude $\sqrt{2\mu\hbar\Delta}$. 
Thus, when the tweezer depths are the same for all states, it is impossible to eject only one of the two particles, and the single particle ejection efficiency is zero.
When we include the tensor Stark shifts of the various molecular states, the trap depths become unequal, which can aid in ejection. 
While the kinetic energy is still shared equally, one of the molecules sees a shallower trap and is ejected.
This results in single-particle ejection efficiencies that reach 100\% within a window of trap depths.
The width of this window is set by the state dependence of the trap depth.
This still holds after averaging over all final states in our case, as shown by the blue curve in Fig.~\ref{fig:ejection}(a).

When the thermal energy is large compared to the difference in trap depths arising from the tensor shifts, it is not \emph{a priori} clear how much of the deterministic ejection survives.
One expects that the difference in trap depth less effectively aids ejection,
but on the other hand the unequal initial momenta can now also lead to single-molecule ejection.
From our MC simulations, we find the window of tweezer depths for which a single molecule is loaded broadens, while the peak performance declines. 
The optimal tweezer depth shifts from around $\hbar\Delta_0/2$ at low temperature, to deeper traps needed to hold on to the molecules with higher thermal energies. 
In this regime, deterministic loading of single molecules is driven by the difference in pre-collision momenta, rather than by the difference in state-dependent trap depth.
In Fig.~\ref{fig:ejection}(b), we highlight the different probabilities of loading zero, one, or two molecules for $T=5$~$\mu$K. 

\subsection{Active ejection\label{sec:active_eject}}

Instead of relying on the thermal energy and differential Stark shifts, we can take a more active approach. One option is to choose a detuning greater than twice the trap depth, so that both molecules can escape, and then rely on the laser cooling to re-capture the molecule that is in $j=1$. The simulations we present below show that the cooling is too slow for this approach to work. A second option is to set the detuning smaller than twice the trap depth so that neither molecule escapes, drive out one of the molecules using a `push beam', and cool the other one. 
A third option is to again set the detuning smaller than twice the trap depth, re-cool the $j=1$ molecule, then briefly lower the trap depth so that the other molecule can escape.

\subsubsection{Push-beam ejection in the purely rotational scheme \label{sec:pushbeamrot}}
Consider the case where the detuning is less than twice the trap depth so that both molecules remain trapped, albeit at a higher energy. After a rotational MWAC, we have $j+j'=1+2$ and the collisional loss rate coefficient will be high due to resonant dipolar interactions. However, the energy release in a MWAC dramatically reduces the density of the two molecules in the tweezer, reducing the rate for the dipolar collisions.
The collision timescale is given by 
\begin{equation}
    \frac{1}{\tau}=k\iiint n_1(\bm{q})n_2(\bm{q})\mathrm{d}\bm{q}\,.
\end{equation}
where $k$ is the loss rate coefficient, $n_i$ is the density distribution of species $i$, and the integral is the density overlap.
The density $n(\bm{q})$ of the barely trapped molecule with mass $m$ at constant energy $E$ in a trap $V(\bm{q})$ is given by the microcanonical ensemble as 
\begin{align}
    n(\bm{q})&=\frac{1}{N}\iiint\delta\left[E-\frac{p^2}{2m}-V\left(\bm{q}\right)\right]\mathrm{d}\bm{p}\nonumber\\
    &=\frac{4\pi m}{N}\sqrt{2m\left[E-V\left(\bm{q}\right)\right]}\,,
\end{align}
with $N$ a normalization constant determined by setting the integral of $n(\bm{q})$ over all space equal to the total number of molecules.

Once the molecule in $j=1$ is cooled to the center of the trap, { which takes about 1~ms \cite{caldwell:19},} the density overlap corresponds well to the peak density of the molecule in $j=2$. For a harmonic trap, this peak density evaluates to $n_\mathrm{harm}(\bm{0})=(\sqrt{2}m^{3/2}\omega_x\omega_y\omega_z/\pi^2)E^{-3/2}$. Before cooling, with both molecules at an energy $E$, the density overlap is a factor $15\pi/32\approx1.47$ smaller than the peak density. For a Gaussian trap, the density is slightly lower, and we evaluate the normalization constant by numerical integration. We assume here a trap with frequencies {$\omega_x=\omega_y=2\pi\times50$~kHz} and {$\omega_z=2\pi\times5$~kHz} with depth $V_0/h=10$~MHz, where we choose {$\Delta=2\pi\times10$~MHz}, such that molecules are excited to $E/h=5$~MHz. With $j=1$ at energies between $E/h=5$~MHz and $E/k_B=5$~$\mu$K during cooling, the overlap density ranges from $3.8\times10^{10}$ to $5.7\times10^{10}$~cm$^{-3}$. Using coupled-channels calculations, we find the dipolar collision rate between molecules in $j=1$ and 2 at a collision energy of 5~MHz to be { $k\sim1\times10^{-9}$~cm$^3$/s}. For the highest density overlap, this gives a collisional timescale of {$\tau\sim20$~ms}, which is long compared to other relevant timescales.

While the $j=1$ molecule is being cooled, the $j=2$ molecule can be ejected by using a resonant laser that pushes it out of the trap. The push is a constant force that tilts the trap potential, making it shallower in the push direction. We use Monte Carlo simulations to model these processes. The initial velocities and positions of the two trapped molecules are drawn at random assuming a thermal distribution at temperature $T$. In the centre-of-mass frame, a collision at time $t=0$ imparts equal and opposite momenta with a random orientation, adding an energy of $\hbar\Delta/2$ to each molecule. We then simulate the trajectories of the two molecules in the tweezer trap. In addition to the restoring force of the trap, there is a damping force for the $j=1$ molecule due to the laser cooling. We take the velocity-dependent force for the `single frequency' deep laser cooling method given in Ref.~\cite{caldwell:19}.  The $j=2$ molecule is pushed by a resonant laser beam that {co-propagates with the tweezer beam and} drives an optical cycling transition ($B ^{2}\Sigma^+(j=1) \leftarrow X ^{2}\Sigma^{+}(j=0,2)$), applying a constant force along the axis of the tweezer. { This is a closed cycle, apart from a leak to $v=1$ with a branching ratio of $2\times 10^{-3}$, which can easily be addressed with a repump laser if needed.} For both molecules, the momentum recoil due to photon absorption and spontaneous emission is included in the model. For $j=1$, we take a mean scattering rate of $2 \times 10^{5}$ photons/s, resulting in an equilibrium temperature of $T=5$~$\mu$K. For $j=2$, the scattering rate is a parameter that we can vary. We simulate $10^4$ molecule pairs. We divide the space into volume elements $\mathrm{d}V$ and count the number of particles in each element to determine  the density distributions $n_{1,2}$ of the two molecules as a function of time, Then we calculate the density overlap $\sum n_1 n_2 \mathrm{d}V$. Multiplying this by $k$ gives the collision rate as a function of time.

\begin{figure}
\centering
\includegraphics[width=\linewidth]{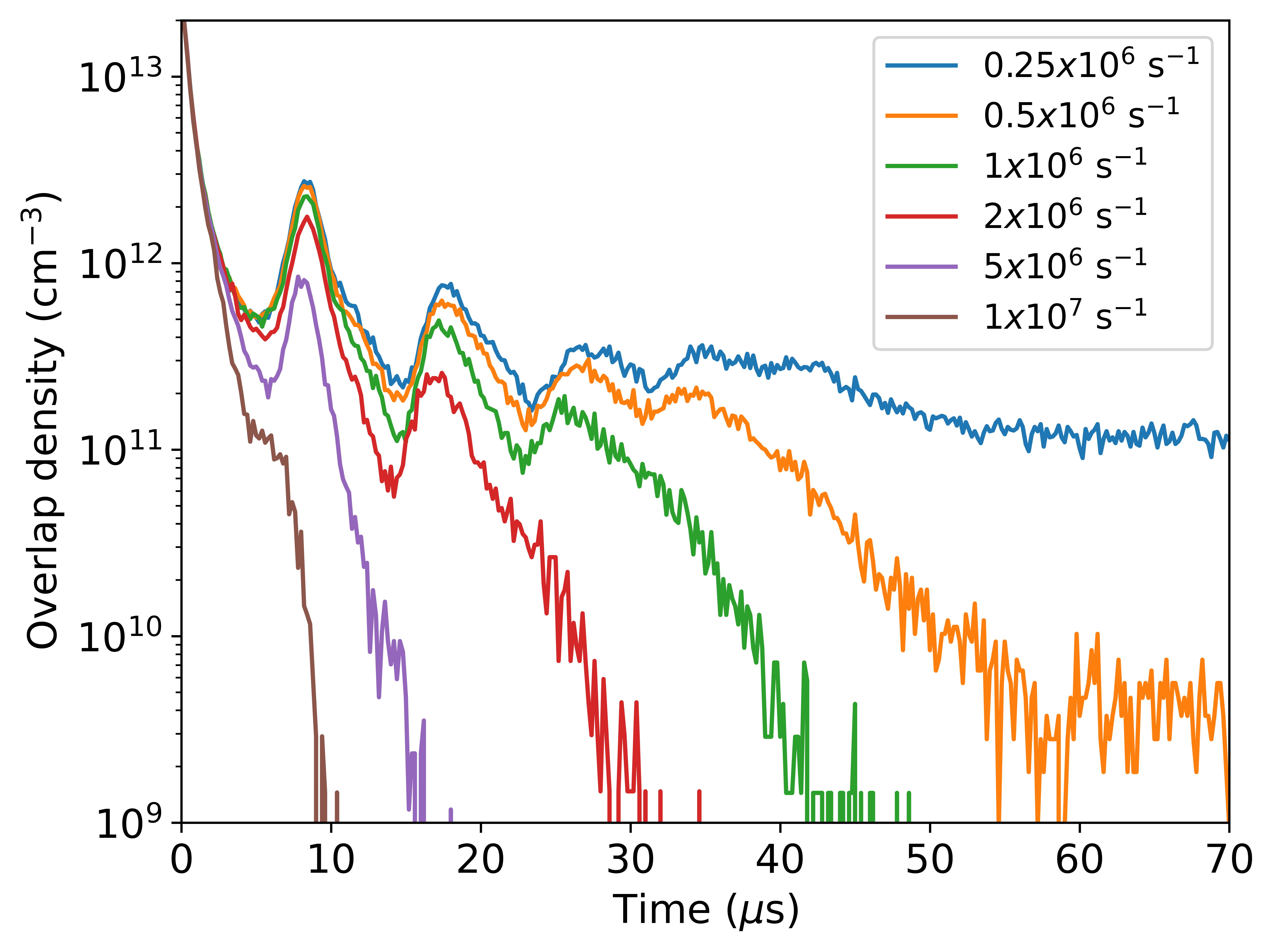}
\caption{
{\bf Ejecting a molecule with a resonant push beam.} Overlap density of a $j=1$ and $j=2$ molecule as a function of time, determined from Monte Carlo simulations. A MWAC occurs at $t=0$, the $j=1$ molecule is re-cooled, and the $j=2$ molecule is pushed out in the axial direction of the tweezer. The legend gives the scattering rate due to the push beam { where we have chosen values that explore the transition from retention to ejection}. The tweezer has a depth of 5~MHz, a wavelength of 780~nm and a waist of 800~nm. The collision energy (detuning) is 0.5 times the trap depth.
\label{fig:overlap_density}
}
\end{figure}

Figure~\ref{fig:overlap_density} shows how the overlap density evolves for six different values of the push force, determined by the scattering rate of the push beam. The overlap density first drops rapidly due to the energy released by the collision, then oscillates due to the radial oscillations in the tweezer, and continues to drop due to the push. When the push is strong (brown line) it overwhelms the confining force of the trap and the particle is rapidly ejected. The overlap density drops below $10^9$~cm$^{-3}$ within 10~$\mu$s in this case. When the push is weak (blue line) the reduction in trap depth due to the tilt may not be sufficient to eject the molecule. Instead, it makes large amplitude oscillations in the trap so there is still a large reduction in density. We see that a scattering rate exceeding $10^6$~s$^{-1}$ is more than sufficient to eject the molecule rapidly. This is 10 times smaller than the maximum possible scattering rate on this transition, confirming the feasibility of this ejection method. Integrating the collision rate over time gives the probability of a dipolar collision. For the results shown in Fig.~\ref{fig:overlap_density}, this probability is 6.0\% for the weakest push, 2.8\% when the scattering rate is $10^6$~s$^{-1}$ and 1.6\% when it has its maximum value of $10^{7}$~s$^{-1}$.. 

\subsubsection{Push-beam ejection in the ro-vibrational scheme \label{sec:pushbeamrv}}

In the ro-vibrational scheme, shelved molecules are in $(v,j)=(2,0)$,
while after the MWAC a pair of molecules is in $(2,0)+(0,1)$. Instead of re-cooling the $j=1$ molecule, we eject it using a push beam resonant with the optical cycling transition $B ^{2}\Sigma^+(j=0) \leftarrow X ^{2}\Sigma^{+}(j=1)$. Note that the pair $(v,j)+(v',j')=(2,0)+(0,1)$ does not experience fast loss by dipolar collisions. {Instead, the molecules experience ro-vibrational van der Waals repulsion, making loss due to collisions in the electronic ground state negligible. In this case, the loss is dominated by collisions between the ground state $(2,0)$ molecules and electronically excited $(0,0)$ molecules. They experience a van der Waals attraction, and we assume that collisions occur at the universal rate of $k_{\rm univ} = 1.4(8) \times 10^{-10}$~cm$^3$/s \cite{cheuk:20}. If we choose the scattering rate to be $10^6$~s$^{-1}$, the excited-state fraction is 2.5\%, and the probability of a collision occurring during ejection is $9.9 \times 10^{-5}$. Thus, in this scheme, the probability of successfully ejecting a single molecule is $P_{\rm eject} = 99.99$\%.}

\subsubsection{Trap-lowering ejection in the ro-vibrational scheme \label{sec:traplowering}}

Consider the case where the detuning is less than twice the trap depth so that after a MWAC both molecules remain trapped.
One of the molecules is in $j=1$ and can be { re-cooled on a 1~ms timescale \cite{caldwell:19}.}
The other molecule is in $j=3$ or $(v,j)=(2,0)$, depending on the chosen scheme, and is not re-cooled.
After this, the tweezer trap depth can be temporarily lowered,
resulting in the ejection of a single molecule.
We emphasize that the asymmetry between the retained and ejected molecule arises not from differential trapping or an initial momentum imbalance,
but purely from the internal state after the MWAC, which determines whether the molecule will be re-cooled.
In tweezers that contain only a single molecule -- either previously shelved or newly loaded -- no MWAC occurs and the molecule remains cold and is not ejected.
Trap-lowering ejection is not recommended for the purely rotational scheme,
where dipolar collisions lead to loss before ejection on the timescale needed for re-cooling of the $j=1$ molecule.
For the ro-vibrational scheme, however, collisional loss is essentially eliminated,
enabling almost deterministic ejection.

\subsection{Enhancing tensor Stark shifts}

Another ejection strategy exploits the different tensor Stark shifts of the rotational states. At low temperatures, we found that a window in tweezer depths exists where deterministic ejection can be 100\%, as was shown in Fig.~\ref{fig:ejection}(a). This window grows with increasing differential Stark shifts. We envision such conditions could be achieved with `anti-magic' tweezers. In contrast to magic conditions where the anisotropic polarizability vanishes \cite{guan:21,gregory:24,ruttley:25}, the wavelength could also be chosen to deliberately increase the difference in polarizability. This asymmetry in tweezer depths could then enable more efficient and deterministic ejection of one of the particles. This requires the difference in tweezer depth to be large compared to $k_BT$. To give an example for CaF, $|\alpha_2| \approx 0.75 |\alpha_0|$ at 650~nm. Raman scattering into other rotational states may become large when $\alpha_2$ is large. At this wavelength, the rate is 1.3~s$^{-1}$ when the trap depth is 5~MHz~\cite{tarbutt:2026}

\section{Shelving}\label{sec:shelve}

Following ejection, each tweezer contains a single molecule, but this can be in one of several states.
In all cases, we wish to bring the molecule back into the cooling cycle, cool it, and finally shelve it.

\subsection{Shelving in the purely rotational scheme}
Depending on the ejection scheme and on whether molecules were loaded in the current and previous cycle,
the remaining molecule can be in one of a number of states.
For example, thermal ejection in the purely rotational scheme retains a molecule in either $j=1$ or $j=2$, or a previously shelved $j=3$ molecule.
These can all be brought to the same state by first transferring collided $(j,f)=(2,3)$ molecules to $(j,f)=(3,3)$ using a microwave pulse that does not affect molecules shelved in $(3,4)$.
Subsequently $j=3$ molecules can be ``de-shelved'' by driving all hyperfine components of $B ^{2}\Sigma^+ (j=2) \leftarrow X ^{2}\Sigma^+ (j=3)$.

The final step is to optically pump these molecules into $(j,f,m_f)=(3,4,4)$.
This can be done by simultaneously driving all hyperfine components of $B ^{2}\Sigma^+ (j=2) \leftarrow X ^{2}\Sigma^+ (j=1)$ and of $B ^{2}\Sigma^+ (j=2) \leftarrow X ^{2}\Sigma^+ (j=3)$. {The former transition can be driven with a beam that co-propagates with the tweezer so that the reservoir is unaffected.} The latter transition { does not affect the reservoir and }should be driven with both $\pi$ and $\sigma^+$ components leaving $(3,4,4)$ as the only dark state in the system.

\subsection{Shelving in the ro-vibrational scheme}

In the ro-vibrational scheme, molecules could be in $(v,j)=(0,1)$, $(0,2)$, or $(2,0).$
These can all be brought to the same state by microwave coupling $j=1\leftrightarrow 2$ in $v=0$ and $j=0\leftrightarrow 1$ in $v=2$,
while applying laser cooling with the $v=1$ re-pump turned off.
The microwave coupling brings all populated rotational states into the cooling cycle.
Turning off the re-pump accumulates molecules in $v=1$ which is unaffected by the microwave dressing, effectively pumping all population into $(v,j)=(1,1).$

After re-cooling these molecules, we wish to shelve them into $(v,j)=(2,0)$. The hyperfine state is less important because the loss suppression and the MWAC efficiency are high for all hyperfine states (see Fig.~\ref{fig:vib_bgloss}(b) and Fig.~\ref{fig:vib_mwac}(a)). There are several ways to reach this state. One way is to turn off one hyperfine component of the $v=2$ repump light used in the laser cooling so that the molecule is pumped into that hyperfine component of $(2,1)$, and then apply a microwave pulse to transfer to $(2,0)$. The molecules in the reservoir are unaffected by the microwave pulse because of the tensor Stark shift, and laser cooling of the reservoir can continue as soon as the shelving is complete. Another way is to first use standard optical pumping within $(0,1)$, then transfer to $(0,0)$ with a microwave pulse, and finally apply a Raman $\pi$-pulse to transfer to $(2,0)$, e.g., via $A ^{2}\Pi_{1/2}(v=1)$.

\section{Iteration\label{sec:performance}}

The sequence of steps is iterated in order to accumulate single molecules in the tweezers.
The filling fraction of tweezers after $n$ cycles, $\varphi_n$, depends on the filling fraction in the previous cycle.
During a new cycle, $k$ molecules are loaded into a tweezer with Poissonian probability $p_k=\lambda^ke^{-\lambda}/k!$,
where $\lambda$ is the loading rate multiplied by the duration of a cycle. 
The overall filling fraction can then be defined recursively as
\begin{align}
    \varphi_n=\;&\varphi_{n-1} \Bigg[p_0 + \Big(\sum_{\mathrm{even}\ k e 2}p_k\Big) + P\Big(\sum_{\mathrm{odd}\ k}p_k\Big)\Bigg]\nonumber\\
    &+(1-\varphi_{n-1}) \Big[\sum_{\mathrm{odd}\ k}p_k\Big]\,.
    \label{eq:poissonmodel}
\end{align}
The last term, proportional to $(1-\varphi_{n-1})$, describes loading of empty tweezers in the collisional blockade regime.
The remaining terms, all proportional to $\varphi_{n-1}$, describe the probability that a loaded tweezer remains loaded.
We distinguish three cases.
If no new molecules are loaded, the shelved molecule remains.
If an even number of molecules is loaded, parity projection removes them rapidly in pairs before the MWAC is initiated at the end of the cycle, and so again the shelved molecule remains.
If an odd number of molecules is loaded, parity projection removes all but one molecule before the MWAC is initiated.
The remaining pair of one shelved and one newly loaded molecule is then converted to a single molecule with efficiency $P$.

In the above, the total single-molecule removal efficiency { $P=P_\mathrm{MWAC}P_\mathrm{eject}P_\mathrm{co-trap} P_\mathrm{spont}$} is the product of four probabilities.
$P_\mathrm{MWAC}$ is the MWAC efficiency discussed in Sec.~\ref{sec:mwac}, and $P_\mathrm{eject}$ is the ejection efficiency discussed in Sec.~\ref{sec:ejection}.
$P_\mathrm{co-trap}$ is the probability that co-trapped molecules survive against collisions, and $P_\mathrm{spont}$ is the probability of remaining in the shelved state without spontaneous emission. Both are discussed below.

The probability that the co-trapped pair of shelved and newly loaded molecules survive the cycle, averaged over the new molecule's arrival time, is given by { $P_\mathrm{co-trap}=\frac{\tau_\mathrm{co-trap}}{t_\mathrm{cycle}} [1-\exp(-t_\mathrm{cycle}/\tau_\mathrm{co-trap})]$ where $\tau_\mathrm{co-trap}$} is the timescale { for collisions between co-trapped shelved and newly loaded molecules}.
{ The} relevant collision rate coefficients are summarized in Table~\ref{tab:rates}.
For the purely rotational scheme, given the { co-trapped} loss rate, a density of $2.4\times10^{13}$~{cm$^{-3}$}, and a 5~ms cycle time, we find { $\tau_\mathrm{co-trap}=2$~ms and $P_\mathrm{co-trap}=37\%$}.
For the ro-vibrational scheme, collisions are so far suppressed that {$\tau_\mathrm{co-trap}$} exceeds seconds.
These collision rates apply to molecules in the electronic ground state,
but laser-cooled molecules also spend part of their time in the electronically excited state.
The excited state fraction is approximately $n_e=2\times 10^{-3}$ for deep laser cooling \cite{caldwell:19}.
In the rotational scheme where the shelved molecule is in $j=3$,
the interaction with the $j=0$ excited state molecule is still of the repulsive rotational van der Waals type,
and the excited state population does not affect the collision rate.
For the ro-vibrational scheme, however, both shelved and electronically excited state molecules are in the rotational ground state,
and they experience an {attractive} rotational van der Waals interaction.
While in the excited state, we can estimate collisions to occur at the universal rate which has been measured to be $k_\mathrm{univ} = 1.4(8) \times 10^{-10}$~cm$^3$/s in the rotational ground state \cite{cheuk:20}.
Despite the small excited state fraction, this dominates the collision rate for co-trapped molecules in the ro-vibrational scheme.
Using this much higher collision rate of $n_e k_\mathrm{univ} = 3 \times 10^{-12}$ cm$^3$/s,
we estimate $P_\mathrm{co-trap}=0.98$ in the ro-vibrational scheme.

Finally, $P_\mathrm{spont}$ is the probability a shelved molecule remains in the shelved state for the duration of a cycle due to the finite lifetime of the shelving state.
In the purely rotational scheme, $P_\mathrm{spont}=1$,
whereas in the ro-vibrational scheme $P_\mathrm{spont} = \exp(-t_\mathrm{cycle}/\tau) \approx 96\%$, given the 120~ms lifetime of the $v=2$ excited state.
If only one of these steps fails, tweezers loaded with an odd number of molecules will be empty,
either because the final MWAC and ejection step failed to convert the co-trapped pair to a single molecule,
or because { collisions between co-trapped molecules} or decay of the shelved state led to parity projection before the MWAC step is even initiated.
When no molecules were loaded, or an even number, none of these failures affect the loading fidelity
because, regardless of whether the shelved molecule is collisionally lost or decays,
a single molecule remains after parity projection, and the subsequent MWAC and ejection steps have no effect.

We note that parity projection in the collisional blockade \cite{schlosser:01,schlosser:02} is assumed above, 
but could be invalid in the presence of genuine three-body loss.
In this case, the recurrence relations can be adapted to reflect a lower-probability loading of single molecules.
In exploratory calculations we have found no effect of this at the low loading rates of 20 molecules per second, assumed throughout this manuscript.

\begin{table}
    \centering
    \caption{\textbf{Collisional rate coefficients}.
    During loading by laser cooling, multiple hyperfine states in $j=1$ are explored, and the { collisional loss rate in each hyperfine state is given}.
    When the microwaves are turned on, MWACs occur at the rate shown.
    In the ro-vibrational shelving scheme, MWACs are induced after state transfer of the newly loaded molecule, so the loading and MWAC stages pertain to different pairs of internal states for the pair of molecules.
    All results are shown for $T=5$~$\mu$K, where MWAC rates are calculated for {$\Delta=2\pi\times10$~MHz} and {$\Omega=2\pi\times1$~MHz}.
    Rate coefficients are given to one digit since the $m_f$ and $m_f'$ dependence affects the rates typically at the level of one to few tens of percents.}
    \begin{tabular}{lcc}
        \hline\hline
        Purely rotational scheme & { Co-trapped} & MWAC rate  \\
        $(j,f)+(j',f')$ & loss rate (cm$^3$/s) & (cm$^3$/s) \\\hline
        \quad\underline{Loading and MWAC} & & \\
        $(1,2)+(3,4)$   & $1\times10^{-11}$ & $2\times10^{-10}$ \\
        $(1,1^+)+(3,4)$ & $2\times10^{-11}$ & $2\times10^{-10}$ \\
        $(1,0)+(3,4)$   & $1\times10^{-11}$ & $1\times10^{-10}$ \\
        $(1,1^-)+(3,4)$ & $1\times10^{-11}$ & $1\times10^{-10}$ \\\hline\hline
        Ro-vibrational scheme & &\\
        $(v,j,f)+(v',j',f')$ & & \\\hline
        \quad\underline{Loading} \\
        $(0,1,2)+(2,0,1)$ &   $1\times10^{-17}$  & - \\
        $(0,1,1^+)+(2,0,1)$ & $2\times10^{-17}$  & - \\
        $(0,1,0)+(2,0,1)$ &   $5\times10^{-17}$  & - \\
        $(0,1,1^-)+(2,0,1)$ & $1\times10^{-17}$  & - \\
        \quad\underline{MWAC} \\
        $(0,2,2^-)+(2,0,1)$ & - &   $7\times10^{-11}$ \\
        \hline\hline
    \end{tabular}
    \label{tab:rates}
\end{table}

\begin{figure}
\centering
\includegraphics[width=\linewidth]{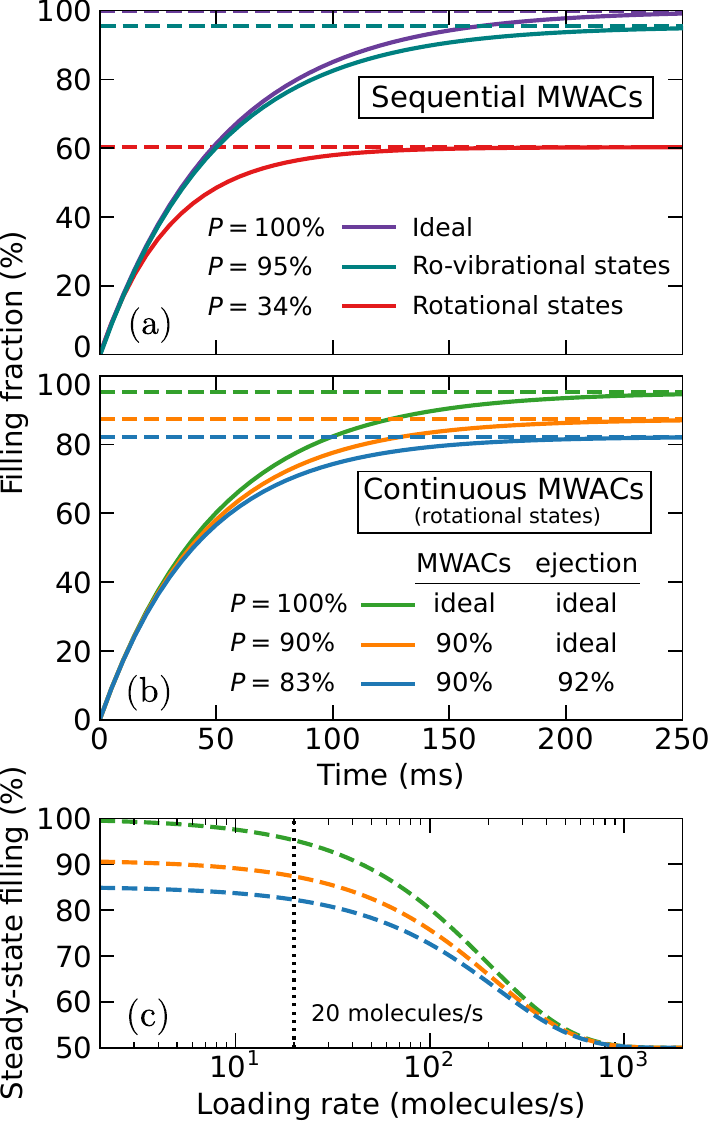}
\caption{
{\bf Loading efficiency of our proposed schemes}. Filling fraction over time given a cycle duration of 5~ms, applying microwave pulses (a) at the end of each loading cycle, or (b) continuously. In all cases, we use {$\Delta=2\pi\times10$~MHz}, {$\Omega=2\pi\times1$~MHz}, $V_0/h=5$~MHz and $T=5$~$\mu$K. The dashed lines represent the steady-state efficiency in the limit of many cycles. Results in (b) are only shown using purely rotational states for a loading rate of 20~molecules/s, which improves the efficiency over sequential MWACs. (c) Dependence of the final steady-state filling fraction on the loading rate, when applying microwaves continuously. The dotted line indicates the loading rate that is illustrated in (b).
\label{fig:performance}
}
\end{figure}

The filling fraction after $n$ cycles can be written in closed form as
\begin{align}
\varphi_n = \frac{ 1-\left[\frac{1}{2}P+\left(1-\frac{1}{2}P\right) \exp(-2\lambda) \right]^n }{2-P}\,.
\label{eq:frac_sequential}
\end{align}
Figure~\ref{fig:performance}(a) shows the filling fraction given by Eq.~(\ref{eq:frac_sequential}) as a function of time, for some relevant values of $P$. The filling fraction increases in time towards the equilibrium value of of $1/(2-P)$. In the case where $P$ is close to unity, this equilibrium is close to $P$.
For the purely rotational scheme, $1/(2-P) = 60\%$, limited primarily by { collisional loss of co-trapped molecules},
and to a lesser extent by the efficiency of the MWAC and thermal ejection steps.
For the ro-vibrational scheme, { $1/(2-P) = 95\%$, limited primarily by the finite lifetime of the $v=2$ vibrationally excited state,
and to a lesser extent by collisions with electronically excited co-trapped molecules during laser-cooling}.
For a loading rate of 20 molecules per { second \cite{schlosser:01,schlosser:02},} this is reached in about 200~ms.
Given that loading of new molecules can operate in the collisional blockade regime, while $(v,j)=(2,0)$ collisional loss is negligible,
one could operate at a higher loading rate~\cite{schlosser:01,schlosser:02},
to reach the same performance even sooner.

Because the performance of the rotational scheme is limited by collisional loss, the scheme can be improved by applying the microwave dressing continuously to induce MWACs as soon as a second molecule is loaded. The ejection of a $j=2$ molecule using a push beam can also be applied continuously. Neither MWAC nor the ejection interfere with the laser cooling. This method eliminates the { collisional loss of co-trapped pairs so that $P_\mathrm{co-trap}=1$}.
A potential drawback is that the MWAC may occur before the newly loaded molecule is cooled completely to the target temperature of $5~\mu$K.
This would reduce the efficiency of thermal ejection, but not the efficiency of pushing $j=2$ molecules out with a resonant laser since this method does not rely on a sharply defined energy release. 

For this continuous MWAC method, the recurrence relation is modified to 
\begin{align}
    \varphi_n=\;&\varphi_{n-1}\Bigg[p_0+(1-P)\Big(\sum_{keq2\ \mathrm{even}}p_k\Big)+P\Big(\sum_{k\ \mathrm{odd}}p_k\Big)\Bigg]\nonumber\\
    &+(1-\varphi_{n-1})\Big[\sum_{k\ \mathrm{odd}}p_k\Big]\,,
    \label{eq:poissonmodel2}
\end{align}
{ with $P=P_\mathrm{MWAC}P_\mathrm{eject}$.
The} main modification is in the term that describes loading of an even number of molecules into a previously loaded tweezer.
Previously, this led to successful loading by parity projection before the MWAC was initiated,
but in this modified scheme the MWAC is initiated as soon as the first new molecule is loaded,
and hence parity projection results in an empty tweezer unless the initial MWAC was unsuccessful, with probability $1-P$.
The filling fraction after $n$ cycles can be written in closed form as
\begin{equation}
    \varphi_n=\frac{\frac{1}{2}(1-e^{-2\lambda})\left\{1-\left[(\left(1-P\right)e^{-2\lambda}+Pe^{-\lambda}\right]^n\right\}}{1-(1-P)e^{-2\lambda}-Pe^{-\lambda}}\,.
    \label{eq:frac_continuous}
\end{equation}
In the case of perfect removal of single molecules, $P=1$, Eq.~(\ref{eq:frac_continuous}) reduces to $\varphi_n=\frac{1}{2}(1+e^{-\lambda})(1-e^{-n\lambda})$ with limit $\varphi_\infty=\frac{1}{2}(1+e^{-\lambda})$,
limited by the finite loading rate. When $P<1$, the largest fraction that can be obtained in the limit of slow loading is $1/(2-P)$, as before. Figure~\ref{fig:performance}(b) shows the filling fraction versus time as given by Eq.~(\ref{eq:frac_continuous}). We see that the steady state filling fraction is higher for this continuous { MWAC} scheme. 
Figure~\ref{fig:performance}(c) shows the steady-state filling fraction versus the loading rate. Lower loading rates yield higher steady-state filling fractions, while requiring more cycles to reach that limit. Loading at 20~molecules/s yields a final fraction of 82\% in the case of thermal ejection and 87\% for active ejection. The limiting factors are the efficiencies of the MWACs and the ejection method.

It is worth considering other potential limitations. The lifetime of the trapped molecules is limited by vibrational heating by blackbody radiation. For CaF in a room temperature environment, the lifetime is 4.5~s. Experiments with laser-cooled molecules often have lifetimes close to this limit e.g.~\cite{jurgilas:2021}, showing that collisions with background gas need not be limiting. Raman scattering into other rotational states has a timescale of 6.3~s when $\lambda=780$~nm and $V_0/h=5$~MHz~\cite{tarbutt:2026}. In laser cooling, there can be loss to higher-lying vibrational states. For laser cooling of CaF it is standard practice to apply vibrational repump lasers up to $v=3$, and for deep laser cooling the lifetime to decay to $v>3$ is about 10~s. These timescales are much longer than the time required to saturate the filling fraction and much longer than the vibrational spontaneous emission timescale. Thus, we conclude that other limitations are small compared to those we have already included. 

\section{Summary\label{sec:conclusion}}

\begin{figure}
\centering
\includegraphics[width=0.8\linewidth]{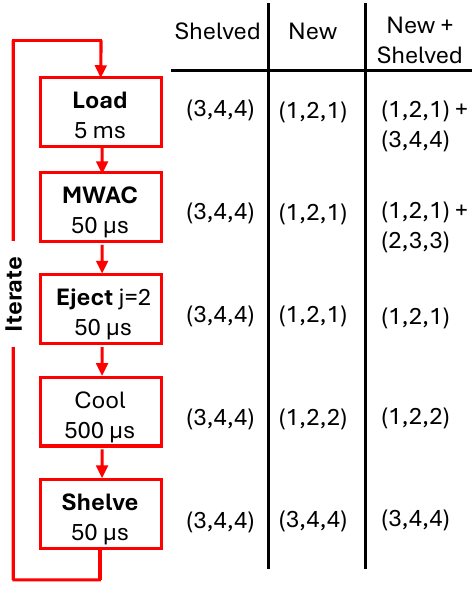}
\caption{
{\bf Summary of recommended scheme using purely rotational states.} Boxes show the steps with indicative timescales, with the 5 main steps highlighted in bold. For ejection, we push the $j=2$ molecule out. The table shows the state(s) $(j,f,m)$ of the molecule(s) at the end of each step, for three cases: a tweezer contains a shelved molecule and no new molecule is loaded; a tweezer does not contain a shelved molecule and a new molecule is loaded; or a tweezer contains a shelved molecule and a new molecule is loaded. Apart from shelving, all steps can be applied continuously.
\label{fig:summary1}
}
\end{figure}

\begin{figure}
\centering
\includegraphics[width=0.8\linewidth]{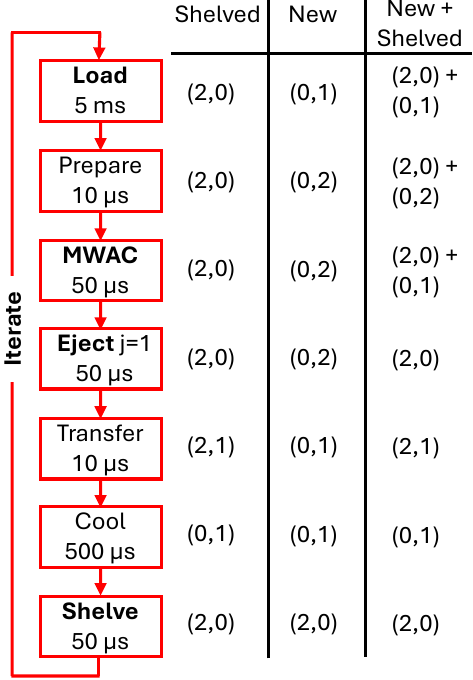}
\caption{
{\bf Summary of recommended scheme using ro-vibrational states.} Boxes show the steps with indicative timescales, with the 5 main steps of the sequence highlighted in bold. For ejection, we push the $j=1$ molecule out. The table shows the state(s) $(v,j)$ of the molecule(s) at the end of each step, for three cases: a tweezer contains a shelved molecule and no new molecule is loaded; a tweezer does not contain a shelved molecule and a new molecule is loaded; or a tweezer contains a shelved molecule and a new molecule is loaded.
\label{fig:summary2}
}
\end{figure}

We have proposed schemes for near-deterministic loading of molecular tweezer arrays using microwave-assisted collisions and efficient ejection.
Microwave-assisted collisions between molecules enable controlled energy release, while keeping short-range collisional loss to a minimum using repulsive van der Waals interactions.
Two schemes are identified, one using the rotational van der Waals interaction \cite{walraven:24a}, the other using a ro-vibrational van der Waals interaction \cite{feng:2026}.
For the subsequent deterministic single-molecule ejection, we have put forward several strategies: thermal ejection, push-beam ejection, trap-lowering ejection, and tensor Stark ejection. 
Thermal ejection utilizes the controlled energy release in a similar way to atomic enhanced loading, but unlike the atomic case is not affected by spontaneous emission since the scheme involves only long-lived rotational states. The initial temperature provides asymmetry in the pre-collision momenta needed to eject a single molecule efficiently.  
In push-beam ejection, the microwave-assisted collision produces molecules with energies below the trap depth so that both are retained. One of the two is then rapidly ejected using a resonant push beam. 
The performance can be limited by collisions that occur between the MWAC and push pulse.
In the rotational scheme this is mitigated by the MWAC energy release which reduces the in-tweezer density,
whereas in the ro-vibrational scheme, the dipolar collisions are eliminated completely.
In trap-lowering ejection, both molecules are retained after the MWAC,
after which one of the molecules is re-cooled and the other ejected by temporarily reducing the tweezer trap depth.
Finally, in tensor Stark ejection, asymmetry in the state-dependent trapping potential is used to deterministically eject single molecules.

Figure~\ref{fig:summary1} summarizes our recommended scheme using purely rotational states, where ejection is achieved by pushing out the $j=2$ molecule. The figure gives indicative timescales for these steps and shows the states of the molecules at the end of each step for the three cases where, after the loading step, the tweezer contains one molecule in $j=3$ (shelved from previous iteration), one molecule in $j=1$ (loaded in current iteration), or a molecule in $j=3$ and a molecule in $j=1$. Apart from the shelving, all steps can be applied continuously, which is more efficient than a sequential scheme, yielding a filling fraction of { up to} 87\%. 

Figure~\ref{fig:summary2} shows our recommended scheme using ro-vibrational states. Here, we choose to push out the $j=1$ molecule and do the steps sequentially. The scheme has more rotational state transfer steps than the pure rotational scheme, but these can be done using microwave $\pi$-pulses or rapid adiabatic passage which have very high fidelities~\cite{holland:25}. All the steps are short compared to the loading time, so the scheme reaches near-deterministic loading without increasing the loading time. The filling fraction of this scheme can be as high as 95\%. 

\section{Outlook\label{sec:outlook}}

The predicted performance of our schemes is comparable to state-of-the-art enhanced loading of atoms, 
which relies on light-assisted collisions that cannot be extended directly to molecules. 
The required experimental tools -- laser cooling, optical pumping, microwave dressing and rotational state transfer -- are all demonstrated. 
The methods increase the filling fraction in single-shot initialization of molecule arrays by more than a factor of two.
More importantly, the improvement in loading efficiency has a disproportionate impact on the assembly fidelity of defect-free arrays.
Increasing the single-tweezer loading probability from stochastic $p \approx 0.35$ to the near-deterministic $p=0.95$ reduces the average defect number $\sim N(1-p)$ by an order of magnitude.
This correspondingly reduces the number of transports required to produce defect-free arrays by rearrangement,
and hence increases the maximum size array that can be assembled without defects with a fixed fidelity by an order of magnitude.

State-of-the-art atom arrays use multi-zone architectures where atoms in a storage zone are continuously replenished from reservoir and preparation zones using an array of auxiliary tweezers~\cite{chiu:25}.
Such architectures will also be highly beneficial for molecule array experiments, especially as losses are more severe for molecules due to the extra loss induced by absorption of blackbody photons. The ideas introduced in this paper could be used to great advantage in such a multi-zone, continuous operation architecture. 

Following the methods demonstrated for atoms~\cite{chiu:25}, we envisage a storage array that is continuously replenished with molecules from a preparation array, that is in turn loaded from a reservoir.
Extraction tweezers repeatedly grab single molecules from the reservoir with probability $q$, and transfer them to traps in the preparation zone. Repeating this multiple times for each tweezer, combined with MWACs to project doubly-loaded tweezers to singly-occupied ones,
loads each tweezer with fidelity $(1-[1-q(2-P)]^m)/(2-P)$,
which can saturate for a modest number of reloads $m$ depending on the extraction probability $q$.
The staturated fidelity is limited by $P=P_\mathrm{MWAC}P_\mathrm{eject}$,
the product of the MWAC and ejection efficiency.
Since the second molecule enters by the extraction tweezer rather than stochastically, the MWAC and ejection can be initiated immediately and the scheme does not require co-trapping of molecules for extended periods. Moreover, laser cooling does not need to be applied in the preparation zone, so the shelving and microwave dressing no longer need to be compatible with laser cooling. This provides more freedom to choose the molecular states that work best for our methods. 

For example, our ro-vibrational scheme can be improved by shelving molecules into $(v,j)=(0,2)$ in the preparation zone, eliminating spontaneous emission of shelved molecules which was previously the limiting factor. Molecules in the extraction tweezers can be cooled and prepared in $(v,j)=(2,0)$, then merged with tweezers in the preparation zone. The MWACs are then initiated by microwave dressing on either $(v,j)=(0,1)\rightarrow (0,2)$ or $(2,0)\rightarrow (2,1)$, followed by push-beam ejection of one of the two molecules. 
Spontaneous emission of a $(2,0)$ molecule is unlikely since the state preparation, transport and merge take only a few ms, and is harmless in any case since it takes the molecule to $(1,1)$ which will be ejected by the push beam.
Having eliminated the leading sources of error -- spontaneous emission and collisions of co-trapped molecules during laser cooling -- the success probability is limited only by the ejection and MWAC efficiencies, which can exceed $p=0.999$ (see Fig.~\ref{fig:vib_mwac}).
At such high filling fraction, it is likely that no rearrangement would be needed at all. Indeed, this method of preparing single molecules has a substantially higher fidelity than the current imaging fidelities for molecules which typically reach about 95\%~\cite{holland:23,holland:23b}. Thus, these ideas offer an exciting and practical route towards near-deterministic loading of molecular tweezers arrays and improved scalability for quantum simulation, sensing, and computing with ultracold molecules.

\begin{acknowledgements}

This work was supported by NWO VIDI (grant ID 10.61686/AKJWK33335).
J.R and M.R.T acknowledge support by EPSRC through grants EP/W00299X/1 and UKRI2226.

\end{acknowledgements}

\bibliography{bib}

\end{document}